\documentstyle[12pt]{article}

\newcommand{\be}{\begin{equation}}
\newcommand{\ee}{\end{equation}}
\newcommand{\bear}{\begin{eqnarray}}
\newcommand{\ear}{\end{eqnarray}}
\newcommand{\e}{\mbox{e}}

\title{
\vspace{2cm}       
Multiloop Feynman Integrals\\
in the Worldline Approach
\footnote{Talk presented by M.G.Schmidt at the Dubna Joint Meeting:
International Seminar on Path Integrals: Theory and Applications,
and 5th International Conference on Path Integrals from meV to MeV, 
Dubna, Russia, 27-31 May 1996.}
}
\author{\large Michael G. Schmidt$^{a)}$, Christian Schubert$^{b)}$\\[3mm]
\em  $^{a)}$ Institut f\"ur Theoretische Physik,
     Universit\"at Heidelberg,\\
\em  Philosophenweg 16,
     D-69120 Heidelberg, Germany\\
\em  $^{b)}$ Institut f\"ur Theoretische Physik,
     Humboldt-Universit\"at zu Berlin,\\
\em  Invalidenstr. 110, D-10115 Berlin, Germany 
}

\date{}

\begin{document}

\maketitle

\begin{abstract}
We explain the concept of worldline Green functions
on classes of multiloop graphs. The QED $\beta$ -- function
and the 2-loop Euler-Heisenberg Lagrangian
are discussed for illustration.
\end{abstract}

\vspace{-13cm}
\begin{flushright}
HD-THEP-96/44\\
HUB-EP-96/29
\end{flushright}

\thispagestyle{empty}

\newpage

\setcounter{page}{1}

{\parindent0em\bf Introduction}\\

The first quantized approach to quantum field theory
(QFTH) involving the worldline path integral of relativistic
particles is well known in the case of propagators and one-loop
effective actions. Still its consequent use
as an alternative to Feynman diagram calculations has been pursued
only recently \cite{1a} (compare our contribution \cite{1} and
references quoted therein): This was triggered by
a systematic discussion of the string tension
$\frac{1}{\alpha'}\to\infty$ limit of string theory \cite{1b}.
The latter is also
usually formulated as a first quantized theory
and it should lead to local QFTH
in this limit. Indeed the relativistic particle quantization on
the worldline was used to argue for a $\sigma$ model formulation
of string theory on the worldsheet. 

Surprisingly the
extension of the worldline formulation to the multiloop case
is much less developed. Starting from the
multiloop string amplitudes \cite{2} and discussing
$\frac{1}{\alpha'}\to\infty$ looks like being the most
straightforward procedure but unfortunately this is technically
very ambitious. One can also use a separate worldline path integral
for each internal propagator \cite{3} or a stringlike
reorganization of standard Feynman parameter integrals \cite{4}. 

We
proposed \cite{5} a multiloop generalization of
Strassler's \cite{1a} approach based on the concept of worldline
Green functions. It is known that in the case of multiloop string
amplitudes there exists a much smaller 
number of topologically different diagrams
per loop order than in ordinary field theory
since the vertices can be deformed continuously.
This is very appealing since it has the prospect of writing
down master expressions for whole classes of Feynman graphs
of a given loop order. The existence of a worldline Green
function characteristic for such a class of Feynman graphs
is essential. Because of the inner vertices these Green
functions are not defined on a one-dimensional manifold (like
for the circle) and hence their existence is nontrivial.
\vskip.3cm

{\parindent0em\bf 2-loop Green functions}\\
\vskip.0cm
We start from the simple 1-loop effective action
in $\phi^3$ theory (mass $m$) which can be written as a
path integral over the circle
(see eqs.(1),(2) of \cite{1}). Correlating two of
the background field insertions at $x(\tau_a),x(\tau_b)$
by a Feynman propagator we obtain a 2-loop expression. The propagator
which we take to be the free one for the beginning can
be written as a Feynman-Schwinger path integral over
$\bar x(\bar\tau)=(x_a(\bar T-\bar\tau)+x_b\bar\tau)/\bar T
+\bar y(\bar\tau)$ with $\bar x(0)=x_a=x(\tau_a)$ and $x(\bar T)
=x_b=x(\tau_b)$ or directly as the well-known propagator in
$x$ -- space
(euclidean dimension $D$)
\\
\be\label{1}
G_F(x_b-x_a)=\int^\infty_0\frac{d\bar T}{(4\pi\bar T)^{D/2}}
\exp\left\{\frac{(x_b-x_a)^2}{4\bar T}-m^2\bar T\right\}.
\ee\\
Then the integral
\vfill\eject

\bear\label{2}
&&\int^\infty_0\frac{dT}{T}\int^\infty_0d\bar T
\frac{1}{(4\pi\bar T)^{D/2}}e^{-m^2(T+\bar T)}
\int^T_0d\tau_ad\tau_b\int[Dx(\tau)]\nonumber\\
&&\times
\exp\left[-\int^T_0d\tau\left(\frac{\dot x^2_\mu}{4}+\frac{(x
(\tau_a)-x(\tau_b))^2_\mu}{4T\bar T}+\lambda\phi(x(\tau))\right)
\right]\ear\\
corresponds to a certain linear combination
of all 2-loop Feynman diagrams with the 
external legs sitting on the loop but not on the inner propagator.

Since $x(\tau_a),x(\tau_b)$ appear quadratically in
the exponential, they can be included in the Gaussian
part of the path integral.
We can obtain a new Green function with the inner line
taken into account by inverting the operator P,
\be\label{3}
\quad
P_{\tau'\tau}=\delta(\tau'-\tau)\frac{\partial_\tau^2}{2}
-\frac{1}{2\bar T}(\delta(\tau'-\tau_a)-\delta(\tau'
-\tau_b))(\delta(\tau_a-\tau)-\delta(\tau_b-\tau))\ee
as\\
\be\label{4}
\quad
G_B^{(1)}(\tau,\tau')=G_B(\tau,\tau')+\frac{1}{2}
\frac{[G_B(\tau,\tau_a)-G_B(\tau,\tau_b)]
[G_B(\tau_a,\tau')-G_B(\tau_b,\tau')]}
{\bar T+G_B(\tau_a,\tau_b)}\ee
where $G_B(\tau,\tau')=|\tau-\tau'|-\frac{(\tau
-\tau')^2}{T}$ is the Green function on the circle
discussed in \cite{1a},\cite{1}. We also obtain a new 
Gaussian path integration determinant,
\\
\be\label{5}
(4\pi T)^{-1/2}[1+\frac{1}{\bar T}G_B(\tau_a,\tau_b)]
^{-D/2}\ee
(with the first factor being the 1-loop determinant). 
Contributions to the 
$n$-point scattering
amplitude are obtained from (\ref{2}) by
expanding for a plane wave background $\phi=\sum_{j=1}^n
e^{ip_jx(\tau)}$. Contraction of the $e^{ip_jx(\tau_j)}$ with
the new Green function (\ref{4}) leads to 
the following integral \cite{5}\\
\bear\label{6}
&&
\int^\infty_0\frac{dT}{T}\int^\infty_0 \!\!d\bar Te^{-m^2
(T+\bar T)}(4\pi)^{-D}\int^T_0\!\!d\tau_a
d\tau_b[T\bar T+TG_B(\tau_a,\tau_b)]^{-D/2}
\nonumber\\
&&\times\prod^n_{i=1}\int^T_0d\tau_i\exp[\sum_{k<l}G_B^{(1)}(\tau
_k,\tau_l)p_kp_l+\frac{1}{2}\sum^n_{k=1}G_B^{(1)}
(\tau_k,\tau_k)p_k^2]\ear\\
We have compared \cite{5},\cite{7} this with 
the ordinary Feynman diagram
parametrization: The $\alpha$ 
parameters correspond to differences
of $\tau$ parameters, and each Feynman graph can be recast into
the form (\ref{6}) with the $\tau$-integration 
restricted to a certain ordered sector.
Thus the new finding is that the Feynman parameter
expressions looking completely different for different Feynman graphs
can be written as parts of one master expression.

If there are also background couplings to the ``inner line'',
this can be taken into account by using a propagator in the
background. We then also have contractions of the inner
variable $\bar y(\bar\tau)$ mentioned above. In the particular case
of $\phi^3$ a symmetric description in all three
propagator lines is preferable and such a Green function,
now valid for contractions all over the $2$-loop graph,
was given by the present authors \cite{5}.

We should emphasize that the worldline
Green functions are not characteristic for a particular
QFTH but just for the topology of the class of graphs. In the
case of more complicated QFTH's like QED worldline vertex
factors have to be included (and contracted). This was
the main issue in \cite{1}.

In the case of multiloop diagrams constructed from
a closed loop by connecting $2m$ points by propagators,
worldline Green functions can be written down along the
same lines \cite{5}. We found\\
\be\label{7}
G_B^{(m)}=G_B(\tau,\tau')+\frac{1}{2}
\sum_{k,l=1}^m[G_B(\tau,\tau_{a_k})-G_B(\tau,
\tau_{b_k})]A_{kl}^{(m)}[G_B(\tau_{a_l},\tau')-
G_B(\tau_{b_l},\tau')]\ee\\
where the $m\times m$ matrix $A^{(m)}$ is defined by
\be\label{8}
(A^{(m)})^{-1}=\bar T-\frac{1}{2}B;\
\bar T_{kl}=\bar T_k\delta_{kl};\ B_{kl}=G_{B_{a_k,a_l}}
-G_{B_{a_k,b_l}}-G_{B_{b_k,a_l}}
+G_{B_{b_k,b_l}}.\ee\\
The determinant factor is 
${\rm Det}(A^{(m)})^{D/2}$. Also the case
with two closed electron loops connected by photons
can be handled easily \cite{8} (see also \cite{9}).

It is nice to see that our expressions (\ref{7}), (\ref{8})
could be confirmed and generalized
recently by Roland and Sato \cite{10}
by an explicit discussion of the limit $\frac{1}{\alpha'}\to\infty$
in string theory, i.e. considering a two-punctured genus
$(m+1)$ Riemann surface. It turns out that the multiloop
worldline Green functions as given above are just the
leading order terms of the string theory Green functions
in this limit.
\vskip.3cm

{\parindent0em\bf Applications: 2- and 3-loop QED $\beta$-function,
2-loop Euler-Heisenberg Lagrangian}\\

The scalar and spinor QED $\beta$-functions in 2 and higher
loop order are interesting objects: Complicated calculations
have simple results (no $\zeta(2n+1)$ terms,  cancellation
of subdivergences).
The 2-loop exercise can be found in \cite{7b}
in one version, but there are more efficient ones \cite{8}.
For this purpose, we introduce a constant external field
${\bf F}$. In the
Fock-Schwinger gauge (see \cite{1}) $A_\mu=\frac{1}{2}y^\rho
F_{\rho\mu}$ such a field may be absorbed into the
worldline Green function on the circle \cite{11},\cite{12}\\
\be\label{9}
{\cal G}_B(\tau_1,\tau_2)=\frac{1}{2(e{\bf F})^2}\left(
\frac{e{\bf F}}{\sin(\e {\bf F}T)}e^{-ie{\bf F}\dot G_{B_{12}}T}+
ie{\bf F}\dot G_{B_{12}}-\frac{1}{T}\right)\ee
The path integral determinant becomes
$(4\pi T)^{-D/2}\exp \{-\frac{1}{2}{\rm tr}
\ln\frac{\sin e{\bf F}T}{e{\bf F}T}\}.$
This allows for a very efficient evaluation of scattering
amplitudes in a constant background field, 
e.g. photon splitting
in a constant magnetic field \cite{13}.

Considering 
the scalar QED $\beta$-function calculation,
to obtain the $\beta$ -- function coefficient
we need to expand the effective action only to second
order in ${\bf F}$. 
To take an internal photon into account,
the denominator of its $x$ -- space propagator
(in Feynman gauge) 
is treated like the scalar propagator, 
and we have almost
the $\phi^3$ case except that one has to contract 
also vertex factors
$\dot x_\mu(\tau_{a,b})$.
The 2-loop Green function
(\ref{4}) is now written with the new ${\cal G}_B(\tau,\tau')$
instead of the original $G_B$, and a denominator $\bar T
-\frac{1}{2}({\cal G}_{aa}+{\cal G}_{bb}-
{\cal G}_{ab}-{\cal G}_{ba})$ instead of
$\bar T+G(\tau_a,\tau_b)$ since there are ``self contractions''.
The latter is also true for the determinant factor (\ref{5}). The
resulting expression can be easily integrated and the
singular part can be extracted. Note that the use of dimensional
regularization is very natural in this approach.

Counterterms still are formed like in conventional QFTH. We
hope to finally embed this in a worldline procedure. Note that the gauge
freedom for the photon propagator is hidden
in a  total $\tau$ -- derivative.

To generalize this from the scalar to the spinor loop,
we have just to substitute superquantities everywhere,
and to
write down also the fermionic analogue of (\ref{9}). It
is satisfying that even the multiloop Green functions (\ref{7})
and the determinant can be generalized
to superquantities \cite{7b}.

For the 3-loop QED $\beta$ -- function (one electron loop:
``quenched'' part) we \cite{14} can use formulas (\ref{7}),
(\ref{8}) for $m=2$. 
Just for an example, the most
basic parameter integral which we have to solve is\\
\be\label{10}
I=\int^\infty_0dT_1dT_2\int^1_0d\tau_ad\tau_bd\tau_cd\tau_d[(T_1+G
_{B_{ab}})(T_2+G_{B_{cd}})-C^2/4)]^{-D/2}\ee\\
with $C=G_{B_{ac}}-G_{B_{ad}}-G_{B_{bc}}+G_{B_{bd}}$
and $\tau_{a,b}$ and $\tau_{c,d}$ the end
positions of the two
inner photons on the electron loop. 

The electron proper time integral
$\int^\infty_0\frac{dT}{T}e^{-m^2T}T^{6-\frac{3}{2}D}=
\Gamma(6-\frac{3}{2}D)m^{3D-12}$
decouples and just gives the overall $\frac{1}{\epsilon}$ pole.
The singular part of (\ref{10}) is easy to split off and to
calculate, the regular part can be evaluated at $D=4$. 
The variable $C$ can
be integrated out trivially leading to
known 2-parameter integrals. 

The old Euler-Heisenberg-Schwinger
formula for the 1-loop effective action in a constant
electromagnetic field $F$ is 
immediately obtained 
from the above 1-loop path integral
determinant. This can be extended to the 
calculation of the 2-loop correction to this
effective action due to
one photon exchange, as
should be already clear
from what we said about the 2-loop QED $\beta$-function calculation.
In this case we keep all orders in $F$ and we are interested
in the finite part. Renormalization can be
performed and we obtain \cite{11} the known result in an
elegant way.

\vfill\eject
\newcommand{\bib}[1]{\bibitem{#1}}

\end{document}